\documentclass[aps,prl,twocolumn,showpacs,superscriptaddress,preprintnumbers]{revtex4}  
\usepackage{graphicx}  
\usepackage{dcolumn}   
\usepackage{bm}        
\usepackage{amssymb}   
\usepackage{amsmath}   
\usepackage{multirow}
\usepackage{color}
\usepackage{units}

\newcommand{\etal}{{\it et al.}}
\newcommand{\pot}{\ensuremath{8.2\mathord{\times}10^{20}}}

\newcommand\numunue{\ensuremath{\nu_{\mu}\, \rightarrow\, \nu_{e}}}

\newcommand\numunutau{\ensuremath{\nu_{\mu}\, \rightarrow\, \nu_{\tau}}}

\newcommand{\numu}{\ensuremath{\nu_{\mu}}}                   
\newcommand{\nue}{\ensuremath{\nu_{e}}}                      
\newcommand{\nutau}{\ensuremath{\nu_{\tau}}}                 
\newcommand{\anumu}{\ensuremath{\overline{\nu}_{\mu}}}       

\begin{document}
\pacs{14.60.Pq, 14.60.Lm, 29.27.-a}


\title{Improved search for muon-neutrino to electron-neutrino oscillations in MINOS}

\newcommand{\Berkeley}{Lawrence Berkeley National Laboratory, Berkeley, California, 94720 USA}
\newcommand{\Cambridge}{Cavendish Laboratory, University of Cambridge, Madingley Road, Cambridge CB3 0HE, United Kingdom}
\newcommand{\FNAL}{Fermi National Accelerator Laboratory, Batavia, Illinois 60510, USA}
\newcommand{\RAL}{Rutherford Appleton Laboratory, Science and Technologies Facilities Council, OX11 0QX, United Kingdom}
\newcommand{\UCL}{Department of Physics and Astronomy, University College London, Gower Street, London WC1E 6BT, United Kingdom}
\newcommand{\Caltech}{Lauritsen Laboratory, California Institute of Technology, Pasadena, California 91125, USA}
\newcommand{\Alabama}{Department of Physics and Astronomy, University of Alabama, Tuscaloosa, Alabama 35487, USA}
\newcommand{\ANL}{Argonne National Laboratory, Argonne, Illinois 60439, USA}
\newcommand{\Athens}{Department of Physics, University of Athens, GR-15771 Athens, Greece}
\newcommand{\NTUAthens}{Department of Physics, National Tech. University of Athens, GR-15780 Athens, Greece}
\newcommand{\Benedictine}{Physics Department, Benedictine University, Lisle, Illinois 60532, USA}
\newcommand{\BNL}{Brookhaven National Laboratory, Upton, New York 11973, USA}
\newcommand{\CdF}{APC -- Universit\'{e} Paris 7 Denis Diderot, 10, rue Alice Domon et L\'{e}onie Duquet, F-75205 Paris Cedex 13, France}
\newcommand{\Cleveland}{Cleveland Clinic, Cleveland, Ohio 44195, USA}
\newcommand{\Delhi}{Department of Physics \& Astrophysics, University of Delhi, Delhi 110007, India}
\newcommand{\GEHealth}{GE Healthcare, Florence South Carolina 29501, USA}
\newcommand{\Harvard}{Department of Physics, Harvard University, Cambridge, Massachusetts 02138, USA}
\newcommand{\HolyCross}{Holy Cross College, Notre Dame, Indiana 46556, USA}
\newcommand{\IIT}{Department of Physics, Illinois Institute of Technology, Chicago, Illinois 60616, USA}
\newcommand{\Iowa}{Department of Physics and Astronomy, Iowa State University, Ames, Iowa 50011 USA}
\newcommand{\Indiana}{Indiana University, Bloomington, Indiana 47405, USA}
\newcommand{\ITEP}{High Energy Experimental Physics Department, ITEP, B. Cheremushkinskaya, 25, 117218 Moscow, Russia}
\newcommand{\JMU}{Physics Department, James Madison University, Harrisonburg, Virginia 22807, USA}
\newcommand{\LASL}{Nuclear Nonproliferation Division, Threat Reduction Directorate, Los Alamos National Laboratory, Los Alamos, New Mexico 87545, USA}
\newcommand{\Lebedev}{Nuclear Physics Department, Lebedev Physical Institute, Leninsky Prospect 53, 119991 Moscow, Russia}
\newcommand{\LLL}{Lawrence Livermore National Laboratory, Livermore, California 94550, USA}
\newcommand{\LosAlamos}{Los Alamos National Laboratory, Los Alamos, New Mexico 87545, USA}
\newcommand{\MIT}{Lincoln Laboratory, Massachusetts Institute of Technology, Lexington, Massachusetts 02420, USA}
\newcommand{\Minnesota}{University of Minnesota, Minneapolis, Minnesota 55455, USA}
\newcommand{\Crookston}{Math, Science and Technology Department, University of Minnesota -- Crookston, Crookston, Minnesota 56716, USA}
\newcommand{\Duluth}{Department of Physics, University of Minnesota -- Duluth, Duluth, Minnesota 55812, USA}
\newcommand{\Ohio}{Center for Cosmology and Astro Particle Physics, Ohio State University, Columbus, Ohio 43210 USA}
\newcommand{\Otterbein}{Otterbein College, Westerville, Ohio 43081, USA}
\newcommand{\Oxford}{Subdepartment of Particle Physics, University of Oxford, Oxford OX1 3RH, United Kingdom}
\newcommand{\PennState}{Department of Physics, Pennsylvania State University, State College, Pennsylvania 16802, USA}
\newcommand{\PennU}{Department of Physics and Astronomy, University of Pennsylvania, Philadelphia, Pennsylvania 19104, USA}
\newcommand{\Pittsburgh}{Department of Physics and Astronomy, University of Pittsburgh, Pittsburgh, Pennsylvania 15260, USA}
\newcommand{\IHEP}{Institute for High Energy Physics, Protvino, Moscow Region RU-140284, Russia}
\newcommand{\Rochester}{Department of Physics and Astronomy, University of Rochester, New York 14627 USA}
\newcommand{\RoyalH}{Physics Department, Royal Holloway, University of London, Egham, Surrey, TW20 0EX, United Kingdom}
\newcommand{\Carolina}{Department of Physics and Astronomy, University of South Carolina, Columbia, South Carolina 29208, USA}
\newcommand{\SLAC}{Stanford Linear Accelerator Center, Stanford, California 94309, USA}
\newcommand{\Stanford}{Department of Physics, Stanford University, Stanford, California 94305, USA}
\newcommand{\StJohnFisher}{Physics Department, St. John Fisher College, Rochester, New York 14618 USA}
\newcommand{\Sussex}{Department of Physics and Astronomy, University of Sussex, Falmer, Brighton BN1 9QH, United Kingdom}
\newcommand{\TexasAM}{Physics Department, Texas A\&M University, College Station, Texas 77843, USA}
\newcommand{\Texas}{Department of Physics, University of Texas at Austin, 1 University Station C1600, Austin, Texas 78712, USA}
\newcommand{\TechX}{Tech-X Corporation, Boulder, Colorado 80303, USA}
\newcommand{\Tufts}{Physics Department, Tufts University, Medford, Massachusetts 02155, USA}
\newcommand{\UNICAMP}{Universidade Estadual de Campinas, IFGW-UNICAMP, CP 6165, 13083-970, Campinas, SP, Brazil}
\newcommand{\UFG}{Instituto de F\'{i}sica, Universidade Federal de Goi\'{a}s, CP 131, 74001-970, Goi\^{a}nia, GO, Brazil}
\newcommand{\USP}{Instituto de F\'{i}sica, Universidade de S\~{a}o Paulo,  CP 66318, 05315-970, S\~{a}o Paulo, SP, Brazil}
\newcommand{\Warsaw}{Department of Physics, University of Warsaw, Ho\.{z}a 69, PL-00-681 Warsaw, Poland}
\newcommand{\Washington}{Physics Department, Western Washington University, Bellingham, Washington 98225, USA}
\newcommand{\WandM}{Department of Physics, College of William \& Mary, Williamsburg, Virginia 23187, USA}
\newcommand{\Wisconsin}{Physics Department, University of Wisconsin, Madison, Wisconsin 53706, USA}
\newcommand{\deceased}{Deceased.}

\affiliation{\ANL}
\affiliation{\Athens}
\affiliation{\BNL}
\affiliation{\Caltech}
\affiliation{\Cambridge}
\affiliation{\UNICAMP}
\affiliation{\FNAL}
\affiliation{\UFG}
\affiliation{\Harvard}
\affiliation{\HolyCross}
\affiliation{\IIT}
\affiliation{\Indiana}
\affiliation{\Iowa}
\affiliation{\UCL}
\affiliation{\Minnesota}
\affiliation{\Duluth}
\affiliation{\Otterbein}
\affiliation{\Oxford}
\affiliation{\Pittsburgh}
\affiliation{\RAL}
\affiliation{\USP}
\affiliation{\Carolina}
\affiliation{\Stanford}
\affiliation{\Sussex}
\affiliation{\TexasAM}
\affiliation{\Texas}
\affiliation{\Tufts}
\affiliation{\Warsaw}
\affiliation{\WandM}

\author{P.~Adamson}
\affiliation{\FNAL}




\author{D.~J.~Auty}
\affiliation{\Sussex}


\author{D.~S.~Ayres}
\affiliation{\ANL}

\author{C.~Backhouse}
\affiliation{\Oxford}




\author{G.~Barr}
\affiliation{\Oxford}






\author{M.~Betancourt}
\affiliation{\Minnesota}



\author{M.~Bishai}
\affiliation{\BNL}

\author{A.~Blake}
\affiliation{\Cambridge}


\author{G.~J.~Bock}
\affiliation{\FNAL}

\author{D.~J.~Boehnlein}
\affiliation{\FNAL}

% ...

\collaboration{The MINOS Collaboration}
\noaffiliation

\date{\today}
\preprint{FERMILAB-PUB-11-351-PPD, BNL-96120-2011-JA}

\begin{abstract}
We report the results of a search for \nue{} appearance in a \numu{} beam in the MINOS long-baseline neutrino experiment.  With an improved analysis and an increased exposure of $8.2\times10^{20}$ protons on the NuMI target at Fermilab, we find that $2\sin^2(\theta_{23})\sin^2(2\theta_{13})<0.12\ (0.20)$ at 90\% confidence level for $\delta\mathord{=}0$ and the normal (inverted) neutrino mass hierarchy, with a best fit of $2\sin^2(\theta_{23})\sin^2(2\theta_{13})\,\mathord{=}\,0.041^{+0.047}_{-0.031}\ (0.079^{+0.071}_{-0.053})$.  The $\theta_{13}\mathord{=}0$ hypothesis is disfavored by the MINOS data at the 89\% confidence level.
\end{abstract}

\vspace*{6pt}
\maketitle

It has been experimentally established that neutrinos undergo flavor change as they propagate~\cite{ref:sol2,ref:reactor,ref:expMuonDis1a,ref:expMuonDis1b,ref:expMuonDis1c,ref:expMuonDis4,ref:expMuonDis5}.  This phenomenon is well-described by three-flavor neutrino oscillations, characterized by the spectrum of neutrino masses together with the elements of the PMNS mixing matrix~\cite{ref:PNMS}.  This matrix is often parametrized by three Euler angles $\theta_{ij}$ and a CP-violating phase $\delta$.  While $\theta_{12}$ and $\theta_{23}$ are known to be large \cite{ref:sol2,ref:expMuonDis1b,ref:expMuonDis4}, $\theta_{13}$ appears to be relatively small~\cite{ref:paloverde,ref:chooz,ref:firstNue,ref:lastNue,ref:global}, with the tightest limits so far coming from the CHOOZ~\cite{ref:chooz} and MINOS~\cite{ref:lastNue} experiments.  The T2K collaboration has recently reported indications of a non-zero value for $\theta_{13}$ at the $2.5\sigma$ confidence level (C.L.)~\cite{ref:T2Knue}.  This letter reports new $\theta_{13}$ constraints from the MINOS experiment, using an increased data set and significant improvements to the analysis.

MINOS is a two-detector long-baseline neutrino oscillation experiment situated along the NuMI neutrino beamline~\cite{ref:numi}.  The 0.98-kton Near Detector (ND) is located on-site at Fermilab, 1.04~km downstream of the NuMI target.  The 5.4-kton Far Detector (FD) is located 735~km downstream in the Soudan Underground Laboratory.  The two detectors have nearly identical designs, each consisting of alternating layers of steel (2.54~cm thick) and plastic scintillator (1~cm).  The scintillator layers are constructed from optically isolated, 4.1~cm wide strips that serve as the active elements of the detectors.  The strips are read out via optical fibers and multi-anode photomultiplier tubes.  Details can be found in Ref.~\cite{ref:minosnim}.

The data used in this analysis come from an exposure of \pot{} protons on the NuMI target.  The corresponding neutrino events in the ND have an energy spectrum that peaks at 3~GeV and a flavor composition of 91.7\% \numu{}, 7.0\% \anumu{}, and 1.3\% $\nu_e\mathord{+}\overline{\nu}_e$, as estimated by beamline and detector Monte Carlo (MC) simulations, with additional constraints from MINOS ND data and external measurements~\cite{ref:expMuonDis4,ref:na49}.   The two-detector arrangement and the relatively small intrinsic $\nu_e$ component make this analysis rather insensitive to beam uncertainties.  Neutrino-nucleus and final-state interactions are simulated using \texttt{NEUGEN3}~\cite{ref:neugen3}, and particle propagation and detector response are simulated with \texttt{GEANT3}~\cite{ref:geant3}.

MINOS is sensitive to $\theta_{13}$ through \numunue{} oscillations.  To leading order, the probability for this oscillation mode is given by
\begin{equation*}\label{equ:oscprob}
P(\numunue)\approx\sin^2(\theta_{23})\sin^2(2\theta_{13})\sin^2(1.27\Delta m_{32}^2 L/E)\ ,
\end{equation*}
where $\Delta m_{32}^2$ (in units of eV$^2$) and $\theta_{23}$ are the dominant atmospheric oscillation parameters, $L$ (in km) is the distance between the neutrino production and detection points, and $E$ (in GeV) is the neutrino energy.  We set constraints on $\theta_{13}$ by searching for an excess of \nue{} events at the FD.  Matter effects and possible leptonic CP violation modify the above probability significantly~\cite{ref:3flav}, hence our results are presented as a function of $\delta$ and the neutrino mass hierarchy.

Events in the MINOS detectors can be characterized by the spatial patterns of energy deposition in the scintillator strips.  Charged-current (CC) \numu{} interactions are identified by a muon track extending beyond the more localized hadronic recoil system.  A single detector plane is 1.4 radiation lengths thick, so the electron from a \nue{} CC interaction penetrates only a few planes (typically 6$-$12), leaving a transversely compact pattern of activity intermingled with the associated hadronic shower.  Neutral-current (NC) interactions can mimic this pattern, particularly when neutral pions are present.

To obtain a \nue{}-enriched sample, we apply a series of selection criteria to the recorded neutrino events.  We require that the neutrino interaction occur within a fiducial volume.  We eliminate most \numu{} CC interactions by rejecting events with a track longer than 24 planes and events with a track extending more than 15 planes beyond the hadronic shower.  We require that an event have at least five contiguous planes with energy greater than half that deposited on average by a minimum ionizing particle.  The calorimetrically determined event energy must lie between 1 and 8~GeV, as events below 1~GeV are overwhelmingly from NC interactions and events above 8~GeV have negligible \numunue{} oscillation probability.  The time and reconstructed direction of each event must be consistent with the low-duty-cycle NuMI neutrino source.  These ``pre-selection'' criteria preserve 77\% of oscillation-induced \nue{} CC events originating in the fiducial volume while passing 8.5\% of \numu{} CC, 39\% of NC, 54\% of \nutau{} CC, and 35\% of intrinsic \nue{} CC events, as estimated by the simulation.


Further background suppression requires a more sophisticated examination of the energy deposition patterns.  Earlier MINOS \nue{} appearance searches used an artificial neural network event classifier with eleven input variables characterizing the transverse and longitudinal profiles of an event's activity in the detector~\cite{ref:firstNue,ref:lastNue,ref:tingjun}.  The present analysis uses a nearest-neighbors algorithm dubbed ``library event matching'' (LEM)~\cite{ref:pedro}.  In LEM, each candidate event is compared to $5\mathord{\times}10^7$ simulated signal and background events~\cite{footnote:LEM}, one by one, to find the 50 that look most similar to the candidate event.  A library event is rejected if its reconstructed energy, number of active strips, or number of active planes differs from that of the candidate event by more than 20\%.  The similarity of the candidate event to each remaining library event is quantified by the following likelihood:
\begin{equation*}\label{equ:lemlogl}
{\log\mathcal{L}} = \sum_{i=1}^{\mathrm{N}_{\mathrm{strips}}} \log \left[ \int_{0}^{\infty} P(n_{\mathrm{cand}}^{i};\lambda)\, P(n_{\mathrm{lib}}^{i}; \lambda)\, d\lambda \right ],
\end{equation*}
where the sum is taken over all strips with a signal above 3~photoelectrons in either of the two events, $n_x^i$ is the charge (in photoelectrons) observed on strip $i$ in event $x$ (with $x$ either ``candidate'' or ``library''), and $P(n;\lambda)$ is the Poisson probability for observing $n$ given mean $\lambda$.  Since events occur throughout the detector volume, each event is translated to a fixed reference location before $\mathcal{L}$ is evaluated.  Strips far away from the event's central axis are combined before comparison.  Additionally, library events are shifted by $\pm1$ plane in search of a better likelihood.

\begin{figure}[hbtp]
\begin{center}
\includegraphics[viewport=0 -20 380 566, keepaspectratio, angle=90,width=0.48 \textwidth]{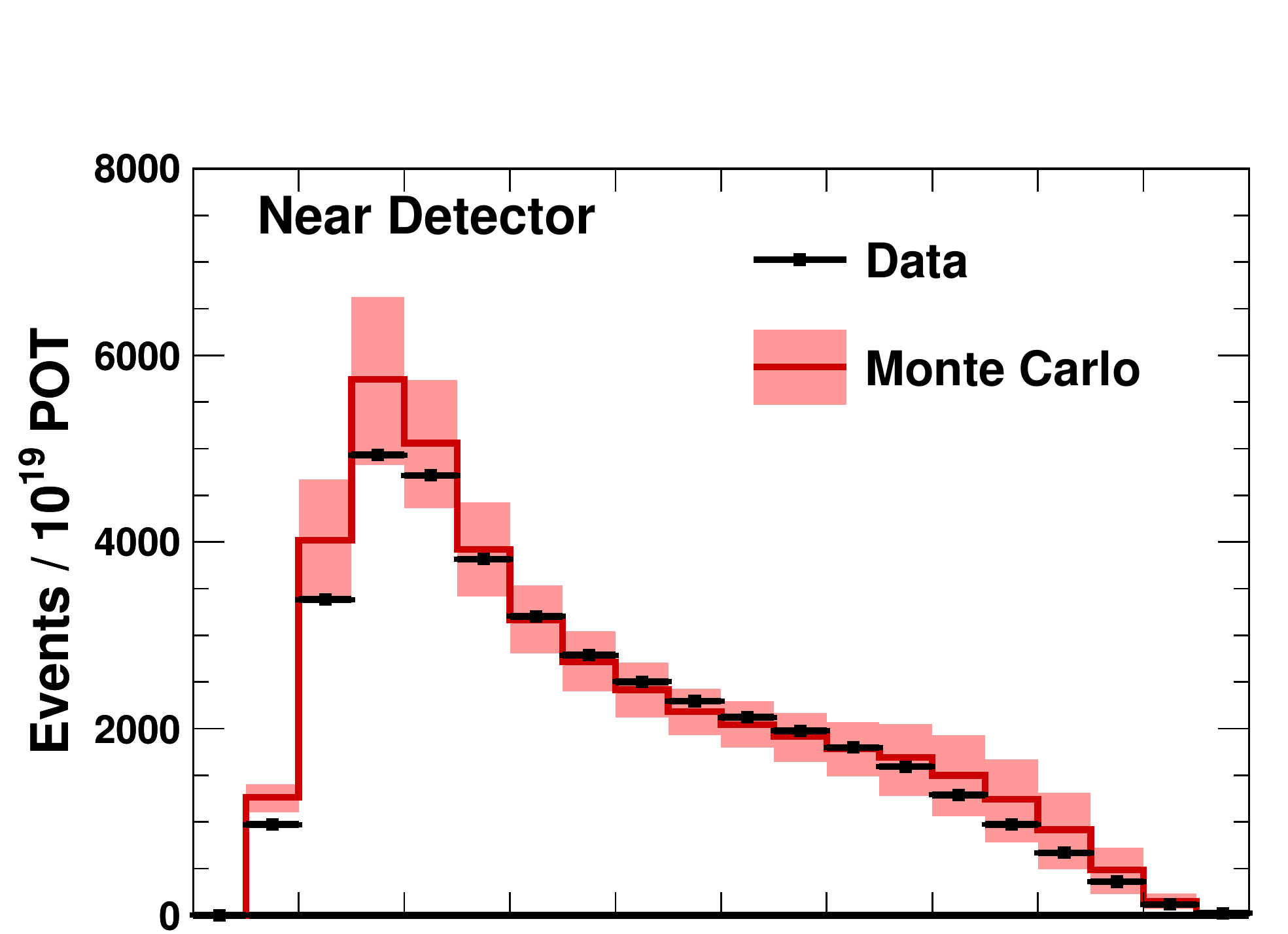}
\includegraphics[viewport=5 -20 398 566, clip=true, keepaspectratio, angle=90,width=0.48 \textwidth]{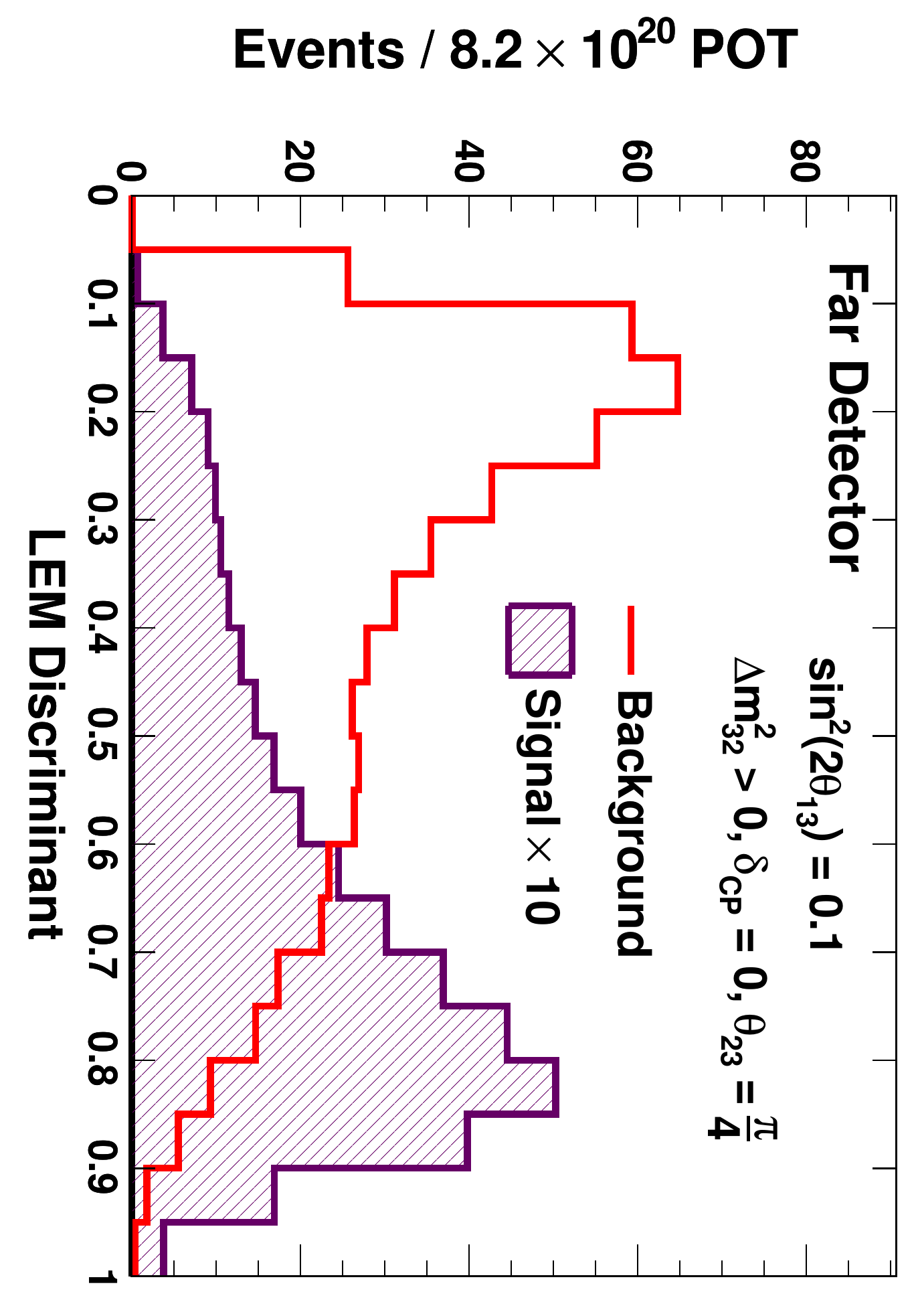}
\end{center}
\caption{[Top] Distribution of the LEM discriminant for events in the Near Detector that pass the pre-selection requirements.  Data (points) and Monte Carlo simulation (histogram) are shown, with the magnitude of the systematic uncertainty indicated by the band.  This uncertainty is highly correlated between the ND and FD and thus cancels out to a large degree when we form our FD predictions.  [Bottom] Expected background and signal distributions in the Far Detector for $\sin^2(2\theta_{13})\mathord{=}0.1$.  The signal distribution has been multiplied by 10 for visibility.}
\label{fig:ndlem}
\end{figure}

The final classifier is formed using a neural network that takes as its inputs the reconstructed event energy along with three variables derived from the best-match ensemble: (1) the fraction of the 50 best-matched events that are true \nue{} CC events, (2) the average inelasticity $\langle y\rangle$ of those \nue{} CC events, and (3) the average fraction of charge that overlaps between the input event and each \nue{} CC event.  The resulting LEM discriminant is shown in Fig.~\ref{fig:ndlem}.

We form a prediction for the FD event rate, in each of 15 bins (specified below) of LEM discriminant and reconstructed energy, using the corresponding rate observed in the ND.  The ND rates are first broken down into individual background contributions, as different background types translate differently from the ND to the FD due to oscillations and beamline geometry.  To determine the relative background contributions in the ND rates, we apply the $\nue{}$ selection to ND data collected in multiple beam configurations with differing neutrino energy spectra and thus differing background compositions.  This allows the construction of a system of linear equations that can be solved for the relative contributions of NC, $\nu_{\mu}$ CC, and intrinsic $\nu_{e}$ CC backgrounds in the primary low-energy beam configuration~\cite{ref:lastNue}.   The measured composition of ND events, averaged over the range LEM$>$0.7 for reconstructed energy between 1 and 8~GeV, is $(61\pm1)$\% NC, $(24\pm1)$\% \numu{} CC, and $(15\pm1)$\% \nue{} CC.

We convert the resulting decomposed ND rates directly into predictions for the FD rates using a Monte Carlo simulation.  More specifically, we use the simulated ratio of FD and ND rates, for each background type and for each LEM and energy bin, as the conversion factor for translating the measured ND rate into the FD prediction.  We evaluate uncertainties on these ratios using systematically modified samples of simulated ND and FD data.  The dominant systematic effects are summarized in Table~\ref{tab:systematics}.
\begin{table}
\begin{tabular}{l c c}
\hline
\hline
\multirow{2}{*}{Uncertainty source} &\ \ & Uncertainty on\\
                                    &\ \ & background events\\
\hline 
Event energy scale           && 4.0\%\\
\nutau{} background          && 2.1\%\\
Relative FD/ND rate          && 1.9\%\\
Hadronic shower model        && 1.1\%\\
All others                   && 2.0\%\\
\hline			  
Total                        && 5.4\%\\
\hline
\hline 
\end{tabular}
\caption{Systematic uncertainties on the number of predicted background events in the FD in the signal region, defined by LEM$>$0.7.  The final $\theta_{13}$ measurement uses multiple LEM and reconstructed energy bins and thus uses a full systematics covariance matrix.  These uncertainties, which are small compared to the statistical errors, lead to a 7.0\% loss in sensitivity to $\sin^2(2\theta_{13})$.  The ``All others'' category includes uncertainties relating to the neutrino flux, cross sections, detector modeling, and background decomposition.}
\label{tab:systematics}
\end{table}

Since the ND collects negligibly few events arising from \numunue{} or \numunutau{} oscillations, the FD rates for these events are estimated using the simulation plus the observed \numu{} CC rates in the ND.  For \nue{} CC events, we further apply an energy- and LEM-dependent correction to the FD predictions that is derived from hybrid events composed of electrons from simulation and hadronic showers from data.  The hadronic showers are obtained by removing the muon hits from cleanly identified \numu{} CC events~\cite{ref:lastNue,ref:josh,ref:anna}, and the electromagnetic shower simulation is verified using a pure sample of electrons recorded by the MINOS Calibration Detector~\cite{ref:caldet}.  The breakdown of expected FD events is given in Table~\ref{tab:fdevents}.  An analysis of beam-off detector activity yielded no \nue{} candidate events, resulting in a 90\% C.L.\ upper limit on cosmogenic backgrounds in the primary analysis region of 0.3 events.  We find that $(40.4\,\mathord{\pm}\,2.8)\%$ of \nue{} CC signal events end up in the signal region, LEM$>$0.7.

\begin{table}[t]
  \begin{center}
\begin{tabular}{cccc}
\hline
\hline
\multirow{2}{*}{Event class} &\ \ & \multicolumn{2}{c}{\ $\sin^2(2\theta_{13})$}\\
                             &\ \ & \ \ \, 0\ \ \, &\ \ \, 0.1\ \ \, \\ \hline
NC                           &\ \ &       34.1   &       34.1        \\
\numu{} CC                   &\ \ &    \ \,6.7   &    \ \,6.7        \\
\nue{} CC                    &\ \ &    \ \,6.4   &    \ \,6.2        \\
$\nu_\tau$ CC                &\ \ &    \ \,2.2   &    \ \,2.1        \\
\numunue{} CC                &\ \ &    \ \,0.2   &       19.1        \\ \hline
Total                        &\ \ &       49.6   &       68.2        \\ \hline
\hline
\end{tabular}
  \caption{Expected FD event counts for $\mathrm{LEM}\mathord{>}0.7$, assuming $\theta_{23}\,\mathord{=}\frac{\pi}{4}$, $\Delta m^2_{32}\,\mathord{=}\,2.32\mathord{\times}10^{-3}\ \mathrm{eV}^2$, and $\delta\,\mathord{=}\,0$.  The first \nue{} line refers to the intrinsic \nue{} component in the beam.  In the $\theta_{13}\mathord{=}0$ case, a small amount of \numunue{} oscillation occurs due to non-zero $\Delta m^2_{21}$.}\label{tab:fdevents}
  \end{center}
\end{table}

Most of the analysis procedures can be tested directly on two signal-free or near-signal-free sideband samples.  First, the ``muon-removed'' hadronic showers described above, before they are merged with simulated electrons, represent a sample of NC-like events.  The predicted and observed LEM distributions in the FD agree for this sample, with $\chi^2/N_{\mathrm{d.o.f.}}\mathord{=}9.7/8$ using statistical errors only.  Second, FD events satisfying $0\mathord{\le}\mathrm{LEM}\mathord{<}0.5$ make up a background-dominated sample for which we predict $370\pm19$ background events (statistical error only).  We observe 377 events, in agreement with prediction.  Forming the prediction for the latter sideband exercises all aspects of the analysis up to the final signal extraction, including the full ND decomposition procedure and the ND-to-FD ratios derived from simulation.

\begin{figure}[hbtp]
\begin{center}
\includegraphics[viewport=5 48 443 718, keepaspectratio,width=0.48 \textwidth]{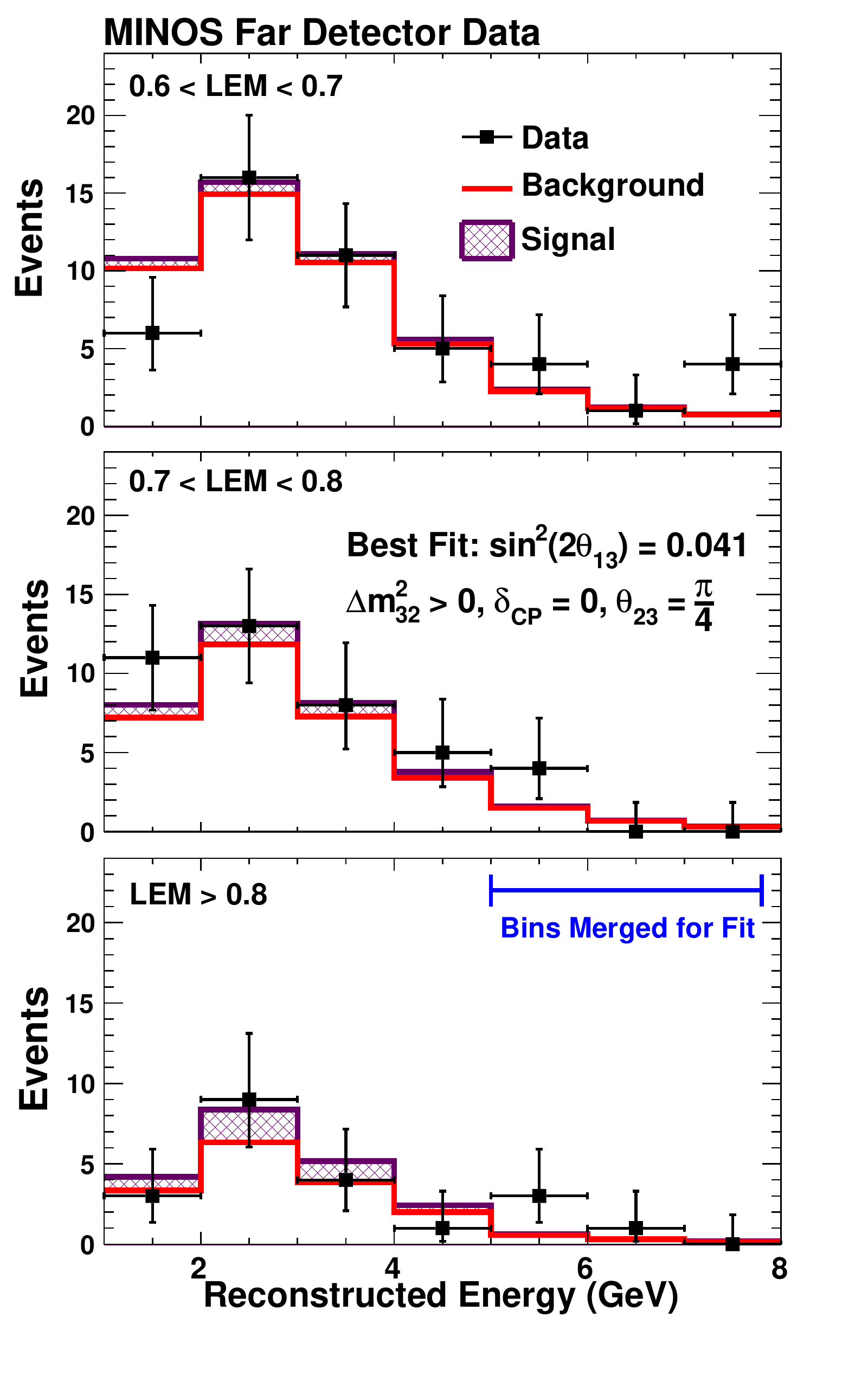}
\end{center}
\caption{Reconstructed energy spectra for \nue{} CC candidate events in the Far Detector.  The black points indicate the data with statistical error bars shown.  The histogram indicates the expected background (unfilled area) together with the contribution of \numunue{} signal (hatched area) for the best-fit value of $\sin^2(2\theta_{13})\,\mathord{=}\,0.041$.}
\label{fig:fddata}
\end{figure}

In previous MINOS analyses~\cite{ref:firstNue,ref:lastNue}, the \nue{} appearance search was conducted by comparing the total number of \nue{} candidate events in the FD to the expected background.  A similar approach applied to the present data yields 62 events in the signal region of $\mathrm{LEM}\mathord{>}0.7$, with an expectation of $49.6\,\mathord{\pm}\,7.0(\mathrm{stat.})\,\mathord{\pm}\,2.7(\mathrm{syst.})$\ if $\theta_{13}\mathord{=}0$.   However, we gain 12\% in sensitivity by fitting the FD sample's LEM and reconstructed energy ($E_{\mathrm{reco}}$) distribution in $3\mathord{\times}5$ bins spanning $\mathrm{LEM}\mathord{>}0.6$ and $1~\mathrm{GeV}\,\mathord{<}\,E_{\mathrm{reco}}\,\mathord{<}\,8~\mathrm{GeV}$.  The energy resolutions for hadronic and electromagnetic showers at 3~GeV are 32\% and 12\%, respectively~\cite{ref:minosnim}.  Figure~\ref{fig:fddata} shows the FD data and predictions used in the fit, along with the extracted best-fit signal.

\begin{figure}[hbtp]
\begin{center}
\includegraphics[viewport=25 40 490 670, keepaspectratio, width=0.48\textwidth]{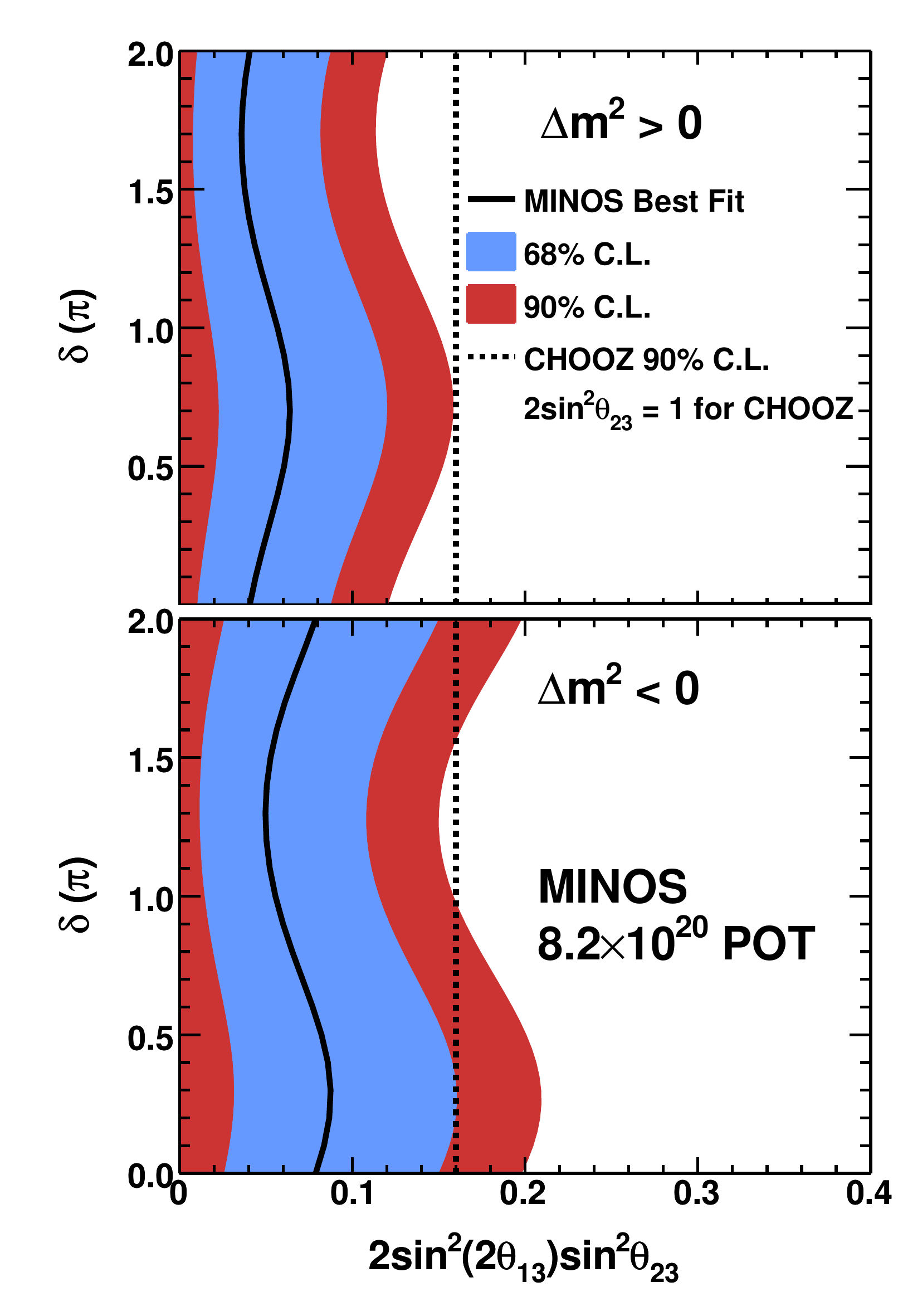}
\end{center}
\caption{Allowed ranges and best fits for $2\sin^2(\theta_{23})\sin^2(2\theta_{13})$ as a function of $\delta$.  The upper (lower) panel assumes the normal (inverted) neutrino mass hierarchy.  The vertical dashed line indicates the CHOOZ 90\% C.L.\ upper limit assuming $\theta_{23}\,\mathord{=}\frac{\pi}{4}$ and $\Delta m^2_{32}\,\mathord{=}\,2.32\mathord{\times}10^{-3}\ \mathrm{eV}^2$~\cite{ref:chooz}.}
\label{fig:contours}
\end{figure}

Figure~\ref{fig:contours} shows the regions of oscillation parameter space allowed by these data.  For the fit, we use a three-flavor oscillation framework~\cite{ref:3flav} including matter effects~\cite{footnote:density}, and we use the Feldman-Cousins procedure~\cite{ref:FC} to calculate the allowed regions.  We assume $\left|\Delta m^2_{\mathrm{32}}\right|\,\mathord{=}\,(2.32^{+0.12}_{-0.08})\mathord{\times} 10^{-3}\ \mathrm{eV}^2$~\cite{ref:expMuonDis4}, $\Delta m^2_{\mathrm{21}}\,\mathord{=}\,(7.59^{+0.19}_{-0.21})\mathord{\times} 10^{-5}\ \mathrm{eV}^2$~\cite{ref:sol2}, $\theta_{23}\,\mathord{=}\,0.785\,\mathord{\pm}\,0.100$~\cite{ref:expMuonDis1b}, and $\theta_{12}\,\mathord{=}\,0.60\,\mathord{\pm}\,0.02$~\cite{ref:sol2}.  The influence of these oscillation parameter uncertainties is included when constructing the contours.

\addtolength{\textheight}{0.27in} 

Prior to unblinding the FD data, we planned to fit only the LEM distribution integrated over energy.  However, the excess over background in the upper energy range prompted the inclusion of energy information so that the fit could weigh events appropriately when extracting $\theta_{13}$ constraints.  If we had performed the signal extraction over LEM bins only, the best fit and 90\% C.L.\ upper limit for $\sin^2(2\theta_{13})$ would each change by +0.006.  A thorough study of high-energy events in the signal and sideband samples, including events between 8 and 12~GeV, indicates that the high-energy predictions are robust and that the selected events are free of irregularities.

In conclusion, using a fit to \nue{} discriminant and reconstructed energy 2D distribution of FD \nue{} candidate events, we find that $2\sin^2(\theta_{23})\sin^2(2\theta_{13})=0.041^{+0.047}_{-0.031}\ (0.079^{+0.071}_{-0.053})$ for the normal (inverted) mass hierarchy and $\delta\mathord{=}0$.  We further find that $2\sin^2(\theta_{23})\sin^2(2\theta_{13})\mathord{<}0.12\ (0.20)$ at 90\% C.L.  Using the less sensitive techniques of the 2010 analysis~\cite{ref:lastNue} on the current data set yields a consistent measurement~\cite{footnote:ann}.  The $\theta_{13}\mathord{=}0$ hypothesis is disfavored by the MINOS data at the 89\% C.L.  This result significantly constrains the $\theta_{13}$ range allowed by the T2K data~\cite{ref:T2Knue} and is the most sensitive measurement of $\theta_{13}$ to date.

This work was supported by the U.S.\ DOE; the U.K.\ STFC; the U.S.\ NSF; the State and University of Minnesota; the University of Athens, Greece; and Brazil's FAPESP, CNPq, and CAPES.  We are grateful to the Minnesota Department of Natural Resources, the crew of the Soudan Underground Laboratory, and the staff of Fermilab for their contributions to this effort.


\begin{thebibliography}{99}

\bibitem{ref:sol2}{B.\ Aharmim \etal{}\ (SNO), Phys.\ Rev.\ Lett.\ {\bf 101}, 111301 (2008).}

\bibitem{ref:reactor}T.\ Araki \etal{}\ (KamLAND), Phys.\ Rev.\ Lett.\ {\bf 94}, 081801 (2005).



\bibitem{ref:expMuonDis1a} Y.\ Ashie \etal{}\ (Super-Kamiokande), Phys.\ Rev.\ Lett.\ {\bf 93}, 101801 (2004).

\bibitem{ref:expMuonDis1b} Y.\ Ashie \etal{} (Super-Kamiokande), Phys.\ Rev.\ D {\bf 71}, 112005 (2005).

\bibitem{ref:expMuonDis1c} K.\ Abe \etal{}\ (Super-Kamiokande), Phys.\ Rev.\ Lett.\ {\bf 97}, 171801 (2006).



\bibitem{ref:expMuonDis4} P.\ Adamson \etal{}\ (MINOS), Phys.\ Rev.\ Lett.\ {\bf 106}, 181801 (2011).

\bibitem{ref:expMuonDis5} P.\ Adamson \etal{}\ (MINOS), Phys.\ Rev.\ Lett.\ {\bf 107}, 021801 (2011).

\bibitem{ref:PNMS}{B.\ Pontecorvo, JETP {\bf 34}, 172 (1958); V.\ N.\ Gribov and B.\ Pontecorvo, Phys.\ Lett.\ B {\bf 28}, 493 (1969); Z.\ Maki, M.\ Nakagawa, and S.\ Sakata, Prog.\ Theor.\ Phys.\ {\bf 28}, 870 (1962).}


\bibitem{ref:paloverde} F.\ Boehm \etal{}\ (Palo Verde), Phys.\ Rev.\ Lett.\ {\bf 84}, 3764 (2000).

\bibitem{ref:chooz}{M.\ Apollonio \etal{}\ (CHOOZ), Eur.\ Phys.\ J.\ C {\bf 27}, 331 (2003).}

\bibitem{ref:firstNue}{P.\ Adamson \etal{}\ (MINOS), Phys.\ Rev.\ Lett.\ {\bf 103}, 261802 (2009).}

\bibitem{ref:lastNue}{P.\ Adamson \etal{}\ (MINOS), Phys.\ Rev.\ D {\bf 82}, 051102 (2010).}

\bibitem{ref:global} G.~L.~Fogli \etal{},\ Phys.\ Rev.\ Lett.\ {\bf 101}, 141801 (2008); M.~C.~Gonzalez-Garcia, M.~Maltoni and J.~Salvado, JHEP {\bf 1004}, 056 (2010); T.~Schwetz, M.~Tortola and J.~W.~F.~Valle, New J.\ Phys.\ {\bf 13}, 063004 (2011).

\bibitem{ref:T2Knue} K.\ Abe \etal{}\ (T2K), Phys.\ Rev.\ Lett.\ {\bf 107}, 041801 (2011).

\bibitem{ref:numi} {K.\ Anderson \etal{},\ FERMILAB-DESIGN-1998-01 (1998).}

\bibitem{ref:minosnim}{D.\ G.\ Michael \etal{},\ Nucl.\ Inst.\ \& Meth.\ A {\bf 596}, 190 (2008).}

\bibitem{ref:na49}C.\ Alt \etal{}\ (NA49),\ Eur.\ Phys.\ J.\ C {\bf 49}, 897 (2007).

\bibitem{ref:neugen3} {S.\ Dytman, H.\ Gallagher, and M.\ Kordosky, arXiv:0806.2119 (2008). }


\bibitem{ref:geant3} {R.\ Brun \etal{}, CERN Program Library W5013 (1984).}

\bibitem{ref:3flav}{E.\ K.\ Akhmedov \etal{},\ J.\ High En.\ Phys.\ {\bf 04}, 078 (2004).}


\bibitem{ref:tingjun} {T.\ Yang, Ph.D.\ Thesis, Stanford Univ.\ (2009).}

\bibitem{ref:pedro} {A description of an earlier version of LEM can be found in J.\ P.\ Ochoa, Ph.D.\ Thesis, Caltech (2009).}

\bibitem{footnote:LEM} {The library composition is 40\% \nue{} CC (signal) and 60\% NC (background).  Only NC background events are needed to obtain good performance, as potentially mis-identified \numu{} CC and \nutau{} CC events look quite similar to NC events.  We note that the performance of LEM is measured on all event classes.}

\bibitem{ref:josh} {J.\ Boehm, Ph.D.\ Thesis, Harvard Univ.\ (2009).}

\bibitem{ref:anna} {A.\ Holin, Ph.D.\ Thesis, Univ.\ College London (2010).}

\bibitem{ref:caldet}{P.\ Adamson \etal{},\ Nucl.\ Inst.\ \& Meth.\ A {\bf 556}, 119 (2006); A.\ Cabrera \etal{}, Nucl.\ Inst.\ \& Meth.\ A {\bf 609}, 106 (2009); P.~Vahle, Ph.D. Thesis, Univ.\ of Texas at Austin (2004).}

\bibitem{footnote:density} {We assume an electron number density of $8.28\mathord{\times}10^{23}\ \mathrm{cm}^{-3}$.}

\bibitem{ref:FC} {G.\ J.\ Feldman and R.\ D.\ Cousins, Phys.\ Rev.\ D {\bf 57}, 3873 (1998).}

\bibitem{footnote:ann} {The older techniques applied to this data set yield a nearly identical best fit value (shifted by $\mathord{+}0.0005$) but a less constraining 90\% C.L.\ upper limit ($\mathord{+}0.015$).}

\end{thebibliography}
\end{document}